\documentclass[aps, prd, twocolumn, lengthcheck, superscriptaddress, showpacs, letterpaper, nofootinbib]{revtex4-1}

\usepackage{graphicx}
\usepackage{color}

\def\la{\langle}\def\ra{\rangle}
\def\be{\begin{eqnarray}}\def\ee{\end{eqnarray}}
\def\lsim{\mathrel{\rlap{\lower3pt\hbox{\hskip1pt$\sim$}}
     \raise1pt\hbox{$<$}}} %less than or approx. symbol
\def\gsim{\mathrel{\rlap{\lower3pt\hbox{\hskip1pt$\sim$}}
     \raise1pt\hbox{$>$}}} %greater than or approx. symbol
\def\le{ \begin{array}{ll}}\def\re{\end{array}}

\def\lear{ \left( \begin{array}{cc}}\def\rear{\end{array} \right)}

\def\le{ \left( \begin{array}{cc}}\def\re{\end{array} \right)}

\def\bi{\bibitem}

\def\del{\partial}

\begin{document}

\title{Pinning Down the Axial-Vector Coupling Constant\\ in Nuclei and Dense Matter}

\author{Mannque Rho}

\affiliation{%
Institut de Physique Th\'eorique,  CEA Saclay, 91191 Gif-sur-Yvette c\'edex, France
}

\date{\today}

\begin{abstract}
The in-medium property of the axial-vector coupling constant $g_A$ in nuclei and dense baryonic matter, a long standing problem in nuclear physics,  is reformulated in terms of the recently constructed scale-invariant hidden local symmetric ($bs$HLS) Lagrangian that encompasses the physics of baryon density ranging from nuclear matter $n_0$ to that of massive compact stars, $\sim 6n_0$. It is shown that unlike the pion decay constant $f_\pi$ that slides with the vacuum change induced by, and dropping with, increasing density, the axial-current constant $g_A$  should remain more or less unaffected by the decrease of chiral condensate in nuclear Gamow-Teller transitions (involving the space component of the axial current)  whereas it gets strongly enhanced  in axial-charge transitions (involving the time component of the axial current)  as density increases.  This phenomenon confirms the ``chiral filter hypothesis" inferred from the current algebras in 1970's. These phenomena can be taken as a ``nuclear physics proof" of Weinberg's folk theorem on quantum effective field theories. The implications of these predictions on giant Gamow-Teller resonances in nuclei and on first-forbidden beta transitions (relevant to nuclear astrophysical processes) are discussed.

\end{abstract}


\maketitle

%%%%%%%%%%%%%%%%%%%%%%%%%%%%%%%%%%%%%%%%%%%%%%%%%%%%%

\section{Introduction}
How the weak axial-vector coupling constant $g_A$ behaves in nuclei has a very long history.  A fundamental quantity, it has impacts on nuclear structure as well as nuclear astrophysical processes. The basic issue raised is associated with how chiral symmetry, a fundamental property of QCD, is manifested in nuclei where the presence of strongly interacting nucleons modifies  the ``vacuum" and hence the quark condensate $\Sigma\equiv |\la\bar{q}q\ra|$, the order parameter of chiral symmetry in QCD, as density is increased. It is now well accepted that the pion decay constant $f_\pi$ should decrease, following the decrease of the condensate $\Sigma$ at increasing density, going to zero (in the chiral limit) at some high critical density $n_c$, which is a fundamental property of QCD.  It seems natural then to expect that the axial coupling constant will undergo a similar {\it intrinsic} decrease in nuclear matter going from  $g_A\approx 1.27$ in the vacuum to $g_A\approx 1$ near the critical density $n_c$.

There has been a suggestion since 1970's~\cite{ wilkinson} that the axial coupling constant, $g_A\approx 1.27$ determined in the matter-free space, quenches to $g_A\approx 1$ in nuclei~\cite{buck-perez}, which has invited an interpretation that it, in consistency with the dropping of $f_\pi$, signals a precursor to chiral restoration. This has led to extensive studies accompanied by controversies in nuclear Gamow-Teller transitions, most notably giant Gamow-Teller resonances~\cite{sakai-suzuki}. The key question asked is whether giant GT resonances signal a quenched axial coupling constant.

In this paper, I revisit this problem and give an extremely simple and unambiguous answer with an effective field theory in which scale symmetry and chiral symmetry of QCD are incorporated. I predict that $g_A$ for Gamow-Teller transitions should remain {\it unaffected} by the quark condensate that slides with density whereas it should get strongly {\it enhanced} in axial-charge (first-forbidden) transitions. This feature, if confirmed, would amount to vindicating the ``chiral filter hypothesis" put forward based on the low-energy theorems of the pre-chiral effective theory era.
\section{Scale-Invariant HLS Lagrangian}
My reasoning exploits the recently formulated effective field theory (EFT) Lagrangian valid for phenomena at low density as well as for high density appropriate for compressed baryonic matter inside such compact stars as the  recently discovered $\sim 2$-solar mass neutron star.

It is constructed by implementing scale symmetry  and hidden local symmetry (HLS) to baryonic chiral Lagrangian consisting of the pseudo-Nambu-Goldstone bosons, pions ($\pi$),  and baryons. The basic assumption that underlies the construction of the effective Lagrangian, denoted $bs$HLS, with $b$ standing for baryons and $s$ standing for the scalar meson, is that there are two hidden symmetries in QCD: one,  scale symmetry broken both explicitly by the QCD trace anomaly and spontaneously with the excitation of a scalar pseudo-Nambu-Goldstone boson, i.e., ``dilaton" $\sigma$;  two, a local flavor symmetry higgsed to give massive $\rho$ and $\omega$.  Neither is visible in QCD in the matter-free vacuum, but the possibility, suggested in \cite{PKLR}, is that both can appear as ``emergent symmetries" in dense matter and control the equation of state (EoS) for highly compressed baryonic matter. I will use this same Lagrangian for calculating nuclear responses to the electro-weak current. There are no unknown parameters in the calculation.

Since the arguments are quite involved and given in great detail elsewhere, I summarize as concisely as possible the essential points that figure in the formulation\footnote{For a recent account related to the issue considered here, see \cite{MR-kps}}. In addition to the nucleon and the pion,   the  Nambu-Goldstone boson of chiral symmetry, there are two additional -- massive -- degrees of freedom essential for the  $bs$HLS Lagrangian: The vector mesons $V=(\rho, \omega)$ and the scalar meson denoted $\chi$. By now very well-known procedure, the vectors $V$ are incorporated by hidden gauge symmetry (HLS)~\cite{HLS}, which is gauge-equivalent to non-linear sigma model at low energy, that elevates the energy scale to the scale of vector mass $\sim 770$ MeV.  The scalar $\chi$ is incorporated~\cite{LMR} by using the ``conformal compensator field" transforming under scale transformation with scale dimension 1, $\chi=f_\chi e^{\sigma/f_\chi}$. Here $\sigma$ is the dilaton field, a pseudo NG boson of scale symmetry,

The underlying approach to nuclear EFT with $bs$HLS is the Landau Fermi-liquid theory based on Wilsonian renromalization group (RG). For this, the ``bare" parameters of the EFT Lagrangian are determined at a ``matching scale" $\Lambda_M$ from which the RG decimation is to be made for quantum theory. The matching is performed with the current correlators between the EFT and QCD, the former at the tree-order and the latter in OPE. The QCD correlators contain, in addition to perturbative quantities, the nonperturbative ones, i.e., the quark condensate $\la\bar{q}q\ra$, the dilaton condensate $\la\chi\ra$, the gluon condensate $\la G_{\mu\nu}^2\ra$ and mixed condensates. The matching renders the ``bare" parameters of the EFT Lagrangian dependent on those condensates.  Since the condensates are characteristic of the vacuum, as the vacuum changes, the condensates slide with the change. Here we are concerned with density, so those condensates must depend on density.  This density dependence, inherited from QCD, is an ``intrinsic" quantity to be distinguished from mundane density dependence coming from baryonic interactions. It is referred to as ``intrinsic density dependence" or  IDD for short.

There are two scales to consider in determining how the IDDs enter in the EFT Lagrangian.

One is the energy scale. The (initial) energy scale is the matching scale from which the initial (or first)  RG decimation is performed.  In principle it could be the chiral scale $\Lambda_\chi\sim 4\pi f_\pi\sim 1$ GeV. In practice it could be lower, say, slightly above the vector meson mass. The scale to which the first decimation is to be made could be taken typically to be the top of the Fermi sea of the baryonic matter.

The other scale is the baryon density.  The density relevant for massive compact stars can reach up to as high as $\sim 6 n_0$. To be able to describe reliably the properties of both normal nuclear matter and  massive stars, a changeover from the known hadronic matter to a different form of matter at a density $\sim 2 n_0$ is required. In  \cite{PKLR}, it is a topology change from a skyrmion matter to a half-skyrmion matter. Being topological this property must be robust. In quark-model approaches, it could be the  hadron-quark continuity that encodes continuous transitions from hadrons to strongly-coupled quark matter or quarkyonic matter~\cite{baymetal}. I believe, as conjectured in \cite{PKLR}, that the two approaches are in some sense equivalent.

The changeover is not a bona-fide phase transition.  However  it impacts importantly on the EoS, making, for instance, the nuclear symmetry energy transform from soft to hard at that density, accommodating the observed $\sim 2$-solar mass star.  Of crucial importance for the process being considered is that when the matter is treated in terms of topological objects, skyrmions, complementary to the $bs$HLS approach, the topology change at $n_{1/2}\sim 2 n_0$ makes the IDDs differ drastically from below to above that density.

It turns out that up to $n \sim n_{1/2}$, the IDD is entirely given by the dilaton condensate $\la\chi\ra$ . The $\chi$ field is a chiral scalar whereas $\bar{q}q$ is the fourth component of the chiral four vector. Therefore the dilaton condensate is not directly connected to the quark condensate, but as mentioned  below, this dilaton condensate gets locked to the pion decay constant which is related to the quark condensate.  While the quark condensate does not figure explicitly in the IDD at low densities,  it controls the behavior of vector-meson masses at compact-star densities, $n\gsim n_{1/2}$~\cite{PKLR}.

 \section{Axial Current with IDD}
 That the IDD could be entirely given by the dilaton condensate was conjectured in 1991~\cite{BR91}, and it has been confirmed to hold up to the density $n\lsim n_{1/2}$~\cite{PKLR}. What is new in the new development is that in the Wilsonian renormalization-group formulation of nuclear effective field theory adopted~\cite{PKLR},  this IDD-scaling is all that figures up to $n_{1/2}$. However it undergoes a drastic change at $n\gsim n_{1/2}$~\cite{PKLR}. This change is important for dense compact-star matter but does not affect the axial-current problem. Nonetheless it constrains properties of lower density matter.

 The effect of the scale-symmetry explicit breaking at the leading order is embedded entirely in the dilaton potential, so it does not enter explicitly
% \footnote{It is of course buried implicitly in the dilaton condensate, which however is determined from the scaling $\Phi$.}
 in the axial response functions in nuclei and nuclear matter that we are interested in. This makes the calculation of the ``intrinsically modified" $g^\ast_A$ in nuclear medium extremely simple.  All we need is the part of the $bs$HLS Lagrangian, scale invariant and hidden local symmetric, that describes the coupling of the nucleon to the external axial field ${\cal A}_\mu$. Writing out explicitly the covariant derivatives involving vector fields, hidden local and external,  and keeping only the external axial vector field ${\cal A}_\mu$, and to the leading order in the explicit scale symmetry breaking, the relevant Lagrangian takes the extremely simple form
\be
 {\cal L} &=&i\overline{N} \gamma^\mu \del\mu N -\frac{\chi}{f_\chi}m_N \overline{N}N +g_A \overline{N}\gamma^\mu\gamma_5  N{\cal A}_{\mu}+\cdots\label{LAG}
 \ee
  Note that the kinetic energy term and the nucleon coupling to the axial field are scale-invariant by themselves and hence do not couple to the conformal compensator field. Put in the nuclear matter background, the bare parameters of the Lagrangian will pick up the medium VeV. Thus in (\ref{LAG}) the nucleon mass parameter will scale while $g_A$ will not:
 \be
 m_N^\ast/m_N=\la\chi\ra^\ast/f_\chi\equiv \Phi, \ \ g_A^\ast/g_A=1\label{Phiscaling}
 \ee
 where $f_\chi$ is the medium-free VeV $\la\chi\ra_0$ and the $\ast$ represents the medium quantities. The first is one of the scaling relations given in \cite{BR91}. The second is new and says that the Lorentz-invariant axial coupling constant {\it does not} scale in density. Now in medium, Lorentz invariance is spontaneously broken, which means that the space component, $g_A^{\rm s}$, could be different from the time component $g_A^{\rm t}$. Writing out the space and time components of the nuclear axial current operators, one obtains
 \be
\vec{J}_A^{\pm} (\vec{x}) &=& g^{\rm s}_A \sum_i \tau_i^{\pm} \vec{\sigma}_i \delta(\vec{x}-\vec{x}_i),\label{GT}\\
 J_{5}^{0\pm} (\vec{x})&=&- g^{\rm t}_A \sum_i\tau_i^\pm  \vec{\sigma}_i \cdot (\vec{p}_i  - \vec{k}/2) /m_N \delta(\vec{x}-\vec{x}_i)\label{axialcharge}
 % + \delta{\vec{x}-\vec{x}_i) + \frac{g^{\rm t}_A}{m_N} \sum_i \tau_i^\pm  \vec{\sigma}_i\cdot \vec{k/2}\delta{\vec{x}-\vec{x}_i) ...
 \ee
 where $\vec{p}$ is the initial momentum of the nucleon making the transition and $\vec{k}$ is the momentum carried by the axial current.  In writing (\ref{GT}) and (\ref{axialcharge}),  the nonrelativistic approximation is made for the nucleon. This approximation is valid not only near $n_0$ but also in the density regime $n\gsim n_{1/2}\sim 2n_0$. This is because the nucleon mass never decreases much after the parity-doubling sets in at $n\sim n_{1/2}$ at which $m_N^\ast\to m_0 \approx (0.6-0.9) m_N$~\cite{PKLR}.

 A simple calculation gives
\be
g_A^{\rm s}=g_A, \ \ g_A^{\rm t}=g_A/\Phi\label{main}
\ee
with $\Phi$ given by (\ref{Phiscaling}).
This is the unequivocal prediction of the IDD with $bs$HLS.

%\vskip 0.2cm
\section{Axial-Charge Transitions}
\begin{figure}[h]
%\vskip -5.cm
%\begin{center}
\includegraphics[width=4.5cm]{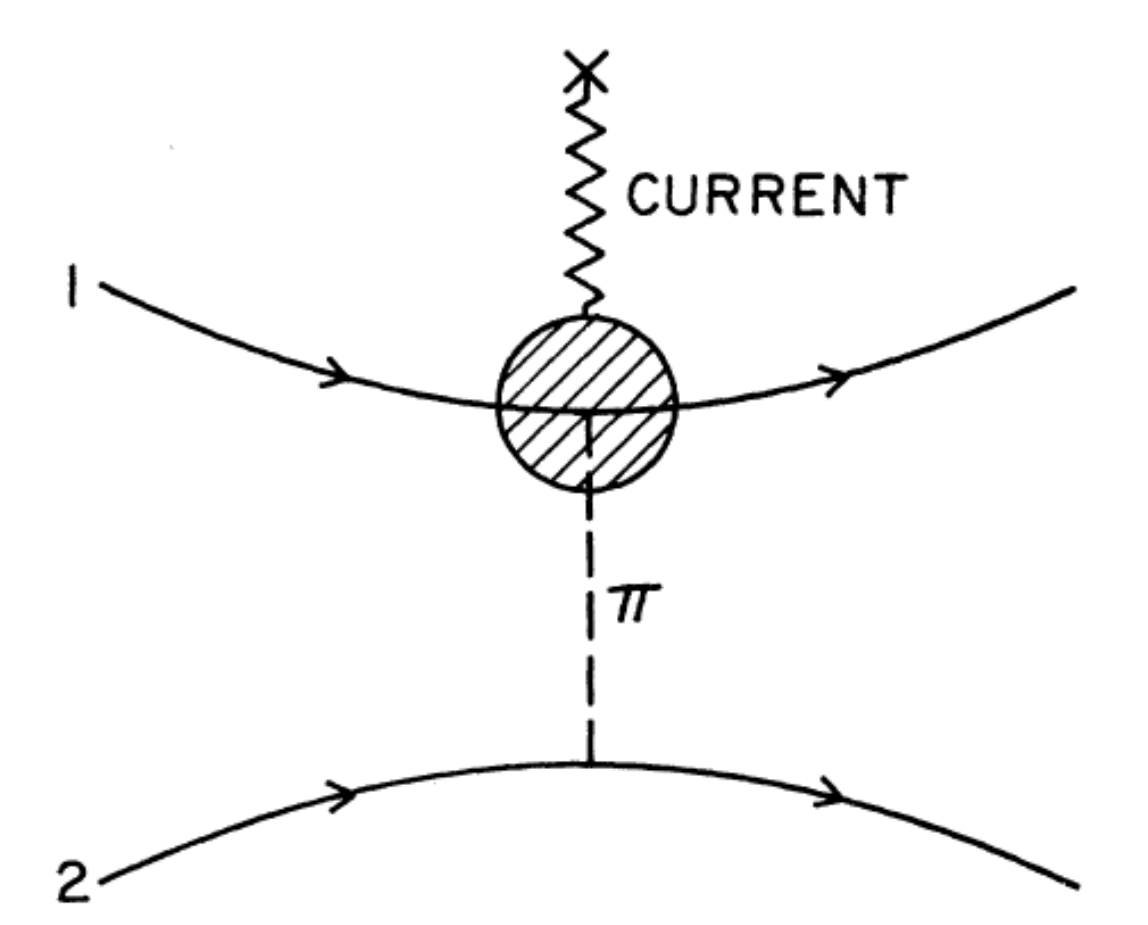}
%\centerline{\epsfig{file=cc.pdf, width=9cm}}
%\end{center}
\caption{Two-body exchange current. The upper vertex involves two soft pions for the axial charge transition.}\label{2-body}
\end{figure}

Let us first consider the transition that involves the time component of the axial current. For this we look at the first forbidden $\beta$-decay process $0^-\leftrightarrow 0^+ $ with $\Delta I=1$ which is governed by the axial charge operator (\ref{axialcharge}). This process has an  axial-vector coupling constant enhanced by $1/\Phi$ (for $\Phi <1$, see below)) and furthermore, more importantly,  receives very important one-pion exchange-current contribution with a vertex ${\cal A}_0 \pi NN$ (in  Fig.~\ref{2-body}). This vertex is of the form of  current algebra with two soft pions and gives an ${\cal O}(1)$ correction to the single-particle operator~\cite{KDR}. The two-body operator is an exactly known pionic-ranged two-body operator, so it can be calculated very accurately if the accurate wave function is known. In fact the ratio $R$ of the two-body matrix element over the one-body matrix element, surprisingly large as a meson exchange-current effect, is highly insensitive to nuclear density. One gets $R=0.5\pm 0.1$ over the wide range of nuclei from light to heavy or in terms of density, $n\sim (0.5-1.0) n_0$~\cite{KR}.
With the two-body effect taken into account,  the effective axial-charge operator is obtained by making the replacement in (\ref{axialcharge})
by
\be
g_A^t\rightarrow g_A^{s\ast} = \epsilon g_A \label{enhancement}
\ee
with
\be
\epsilon=\Phi^{-1} (1+R/\Phi). \label{epsilon}
\ee
To make an estimate of $\epsilon$, we need the scaling factor $\Phi$. It is an easy calculation to find that in medium,  the dilaton condensate is locked to  the quark condensate, leading to~\cite{PKLR}
\be
\Phi=f_\chi^\ast/f_\chi\approx f_\pi^\ast/f_\pi.
\ee
The pion decay constant $f^\ast_\pi$ -- which is a parameter in the theory -- can be extracted from deeply bound pionic nuclear systems~\cite{yamazaki}. It can be parameterized in density (up to $n_0$) as
\be
\Phi\approx 1/(1+0.25 n/n_0).\label{Phi}
\ee
The prediction (\ref{epsilon})  has been well confirmed in experiments ranging from $A=12$ to $A=205-212$~\cite{e-MEC,warburton}. Here for illustrative purpose, let me quote the result in Pb nuclei. Taking the density to be near the nuclear matter density, one has $\Phi (n_0)\approx 0.8$ in accordance with Eq.~(\ref{Phi}) -- which is fit to the experiment~\cite{yamazaki}. Substituting $R=0.5\pm 0.1$, one predicts a huge enhancement factor
\be
\epsilon (n_0)\approx 2.0\pm 0.2.
\ee
This factor  compares well with Warburton's result in Pb region~\cite{warburton}
\be
\epsilon^{exp}=2.01\pm 0.05.
\ee
This result has two important implications. One is that this is the most spectacular effect of the IDD combined with the soft-pion dominance in nuclear axial charge processes. The corrections to the soft-pion contribution in the meson exchange charge operators are highly suppressed, so the transition operator is very precisely determined due to a mechanism referred to in \cite{KDR} as ``chiral filter."  Given that nuclear structure techniques are vastly improved in the two decades and half since 1991, the ratio $R$ could now be  calculable extremely accurately for the ranges of nuclei involved, in particular in the $A=12-16$ region. This enhanced process would constitute the most spectacular and pristine evidence for meson-exchange currents in nuclei.

The second is that this huge enhancement  could have an important impact in astrophysical processes.  For instance it was observed~\cite{honmaetal} that in $N=126$ isotones, the first-forbidden transition -- which is relevant to astrophysics -- affects appreciably the half-lives for larger $Z$: For  $Z=72$, the half-life is quenched by a factor of $\sim 5$ by the  FF contribution. This calculation however does not take into account the $\epsilon$ factor. If the FF process is dominated by the single-particle axial charge operator in this process, a back-of-envelope estimate shows that the soft-pion-exchange contribution could make the half-life an order of magnitude shorter than the estimate given in  \cite{honmaetal}.

\section{Gamow-Teller Transition}
Now let us turn to the Gamow-Teller transition which is dominated by the Gamow-Teller operator (\ref{GT}). The prediction says that there be no renormalization in the Gamow-Teller matrix elements due to partial restoration of chiral symmetry in nuclear medium, at least up to density $n_{1/2}\sim 2n_0$.  In order to confront this prediction with the experimental value, one would have to compute the possible corrections coming from meson-exchange currents.  For this, the ``chiral filter hypothesis" that leads to the accurate axial charge operator dominated by the soft-pions~\cite{KDR,MR91} says the exchange current to the Gamow-Teller operator will be screened by chiral filter and meson-exchange currents will come at higher chiral order, hence unimportant. This means the Gamow-Teller operator will {\it also} be relatively little disturbed by corrections in the operators, given that the axial coupling constant is not modified by the vacuum change driven by density. Given accurate wave functions, the Gamow-Teller transition could also be calculated accurately in consistecy with QCD. A recent {\it ab initio} quantum Monte Carlo calculation in $A=6-10$ nuclei seems to give a support to this prediction~\cite{pastoreetal}.

\section{Problem}
In the renormalization-group (RG) procedure put forward in \cite{PKLR}, although the IDD -- brought in by matching to QCD at the matching scale -- does not affect the $g_A^s$, one has to consider what other degrees of freedom can contribute in the RG decimations. The first energy scale one encounters in the Gamow-Teller channel as one makes the first decimation  from the matching scale is the $\Delta$-hole excitation scale with $E_{\Delta-h}\sim 300$ MeV above the Fermi sea.  Other channels could be safely ignored. From a general hadronic interaction point of view, there is a good reason to believe that  the Gamow-Teller operator would couple importantly to $\Delta$-hole states. When integrated out, the $\Delta$-hole degrees of freedom should modify the axial vector current. Now since the weak current acts only once, this effect can be included in the modification of the Gamow-Teller coupling constant $g_A^s$. This  was worked out a long time ago, which could be phrased in terms of Landau-Migdal's ${g_0^\prime}$ parameter in the $\Delta$-N channel~\cite{Delta,BR-GT}  when treated in Landau's Fermi liquid theory extended to the space of $N$ and $\Delta$. This can also be phrased in terms of the Ericson-Ericson-Lorenz-Lorentz (EELL) effect in pion-nuclear interactions~\cite{EELL}.  If one takes ${g_0^\prime}$ equal in the $NN$, $N\Delta$ and $\Delta$-$\Delta$ channels, that is, universal, then one finds that the effective $g^\ast_A$ for Gamow-Teller transitions in nuclear matter is renormalized to $g_A^\ast\approx 1$.  Note that this is {\it not} associated with the vacuum change due to density. It is a class of nuclear correlation effects at the scale of $\sim 300$ MeV. If one were to compute in the simple shell-model in a single model space, the effective $g^\ast_A$ needed there would be ``quenched" and consistent with what was arrived at in light nuclei~\cite{buck-perez}. Why it should be $g_A^\ast\approx 1$, not some other value, is not clear. If one were to do the calculation in a large model space including higher correlations as in \cite{pastoreetal}, then the effective $g^\ast_A$ could be different, approaching near the effective field theory prediction (\ref{main}). This means that if one were to extend the model space that covers the excitation energy comparable to $\Delta$-hole excitations, then the Gamow-Teller sum rule should be satisfied with $g_A=1.27$ as given by (\ref{main}), that is, without quenching. It seems plausible that the presently accurately measured sum rule in $^{56}$Ni~\cite{sakai-gt} is satisfied with the quenched constant because the strengths at much higher energy than what's covered in the experiment are  missing in the sum rule.

%In order to understand what this ``renormalization" means, imagine doing a large-scale, or preferably  full-scale, shell-model calculation within the nucleon configuration space only. This corresponds to doing the RG decimation from the energy scale $E_{\Delta-h}\approx 300$ MeV. Hence using this {\it renormalized} $g_A^\star\approx 1$ in calculating the GT matrix element, one would find $(g_A^\ast/g_A)^2 \approx 0.6$ quenching in the strength. Indeed an observation of the famous  $\sim 40\%$ quenching in Gamow-Teller strength has baffled experimentalists for many years. However,  more recent experiments on giant Gamow-Teller resonances find that this quenching has more or less disappeared~\cite{sakai-suzuki}. In terms of $g_0^\prime$, this means that $g_0^\prime|_{\Delta N}\ll g_0^\prime|_{NN}$.

It seems plausible that high-order correlations in the nuclear wave functions will reduce the $\sim 20\%$ quenching in $g_A$ found in simple shell-model calculations as demonstrated in \cite{pastoreetal}. The quenching may largely disappear as the excitation reaches the $\Delta$-hole energy. This would confirm the prediction (\ref{main}). {\it  However theoretically this is unnatural and raises a nontrivial question.} This would mean that $g_0^\prime|_{N\Delta}$ must be much smaller than $g_0^\prime|_{NN}$.  It is a well-known fact that $g_0^\prime|_{NN}$ is by far the strongest quasiparticle interaction controlling spin-isospin excitations in nuclei. So why the $\Delta N$ contribution to $g_0^\prime$  is so suppressed compared with the $NN$ channel is difficult to understand; second, at least a part of the Landau parameters can be associated with an IDD. Take for example the in-medium $\rho$ mass with its IDD given by the dilaton condensate $\Phi$.  In the mean-field treatment of the anomalous orbital gyromagnetic ratio of the proton $\delta g_l$ in heavy nuclei~\cite{friman-rho},  it has been shown that the Landau parameter $F_1^\prime$ is also related to $\Phi$ when treated in the single-decimation RG using the $bs$HLS Lagrangian. Thus at least a part of the Landau parameters can be associated with IDD. Which part of the Landau parameters is concerned depends on the precise way the RG decimations are defined.

%The message from what we have learned is this: It is a fundamental task to pin down what the $\Delta$-hole modification of $g_A$ is by doing a precision calculation of giant Gamow-Teller resonances in heavy nuclei and determine the deviation of $g_A$, if any,  from the free-space value. There is no known reason why the deviation should be zero even though  as shown in this paper there is no {\it intrinsic} QCD correction {\it directly associated} with the vacuum change in the chiral condensate.
%

Another interesting issue is the role of NG bosons in nuclear physics. When the energy scale probed in nuclear processes is much less than the pion mass $\sim140$ MeV, the pion can also be integrated out  and one obtains ``pionless effective field theory" for nuclear physics, which is eminently a respectable effective field theory.   With no pions present, however, there is no explicit foot-print of chiral symmetry, i.e., no smoking gun for the spontaneous breaking of chiral symmetry. Yet the theory seems to work fairly successfully in various low-energy processes involving light nuclei including the solar proton fusion process that is dominated by the Gamow-Teller operator. The question is: What about the double-soft process that makes such a huge effect in the axial-charge transitions?  Could it be  hidden in the pionless effective field theory, somewhat like the hidden symmetries discussed above?

Finally I should point out that the present prediction by the modern scale-invariant hidden local symmetry Lagrangian and the old ``chiral filter hypothesis" based on the soft-pion theorems of current algebras provide a nuclear physics proof of Weinberg's ``Folk Theorem" on quantum effective field theories~\cite{weinberg}.

\end{document}